
\input PHYZZX
\title{\bf
Correlation between Instantons and QCD-monopoles
\nextline
in the Abelian Gauge
}

\author{H.~Suganuma$^{\rm a}$, K~Itakura$^{\rm b}$,
H.~Toki$^{\rm a}$ and O.~Miyamura$^{\rm c}$}

\address{
{\rm a)} Research Center for Nuclear Physics (RCNP),
Osaka University,
\nextline
Mihogaoka 10-1, Ibaraki 567, Japan
}

\address{
{\rm b)} College of Arts and Sciences,
Tokyo University,
\nextline
Komaba 3-8-1, Meguro, Tokyo 153, Japan
}

\address{
{\rm c)} Department of Physics, Faculty of Science,
Hiroshima University,
\nextline
Kagamiyama 1-3, Higashi-Hiroshima 739, Japan
}

\abstract{
The correlation between instantons and QCD-monopoles is studied
both in the lattice gauge theory and in the continuum theory.
{}From a simple topological consideration,
instantons are expected to live only around the QCD-monopole
trajectory in the abelian gauge.
First, the instanton solution is analytically studied
in the Polyakov-like gauge, where $A_4(x)$ is diagonalized.
The world line of the QCD-monopole is found to be penetrate
the center of each instanton inevitably.
For the single-instanton solution,
the QCD-monopole trajectory becomes a simple straight line.
On the other hand, in the multi-instanton system,
the QCD-monopole trajectory often has complicated topology
including a loop or a folded structure, and is unstable against
a small fluctuation of the location and the size of instantons.
We also study the thermal instanton system in the Polyakov-like gauge.
At the high-temperature limit, the monopole trajectory becomes
straight lines in the temporal direction.
The topology of the QCD-monopole trajectory is drastically changed
at a high temperature.
Second, the correlation between instantons and QCD-monopoles is studied
in the maximally abelian (MA) gauge and/or the Polyakov gauge
using the SU(2) lattice with $16^4$.
The abelian link variable $u_\mu (s)$ is decomposed into
the singular (monopole-dominating) part $u_\mu ^{Ds}(s)$ and the regular
(photon-dominating) part $u_\mu ^{Ph}(s)$.
The instanton numbers, $Q({\rm Ds})$ and $Q({\rm Ph})$, are measured
using the SU(2) variables, $U_\mu ^{Ds}(s)$ and $U_\mu ^{Ph}(s)$, which
are reconstructed by multiplying the off-diagonal matter factor to
$u_\mu ^{Ds}(s)$ and $u_\mu ^{Ph}(s)$, respectively.
A strong correlation is found between $Q({\rm Ds})$ in the singular
part and the ordinary topological charge $Q({\rm SU(2)})$
even after the Cabibbo-Marinari cooling.
On the other hand, $Q({\rm Ph})$ quickly vanishes by several
cooling sweeps, which means the absence of instantons in
the regular part.
Such a monopole dominance for the topological charge is found
both in the MA gauge and in the Polyakov gauge.
}

\chapter{Topological Consideration}

Recently, essential roles of monopole condensation
\REF\nambu{
Y.~Nambu, Phys.~Rev.~{\bf D10} (1974) 4262.
}
\REF\thooftA{
G.~'t~Hooft, {\it High Energy Physics},
(Editorice Compositori, Bologna, 1975).
}
\REF\mandelstam{
S.~Mandelstam,  Phys.~Rep.~{\bf C23} (1976) 245.
}
\REF\thooftB{
G.~'t~Hooft, Nucl.~Phys.~{\bf B190} (1981) 455.
}
[\nambu-\thooftB]
to the nonperturbative QCD are strongly suggested
by the studies based on the lattice gauge theory
\REF\kronfeld{
A.S.~Kronfeld, G.~Schierholz and U.-J.~Wiese,
Nucl.~Phys.~{\bf B293} (1987) 461.
}
\REF\suzuki{
T.~Suzuki and I.~Yotsuyanagi,
Phys.~Rev.~{\bf D42} (1990) 4257.
}
\REF\hioki{
S.~Hioki, S.~Kitahara, S.~Kiura, Y.~Matsubara,
O.~Miyamura, S.~Ohno and
\nextline
T.~Suzuki, Phys.~Lett.~{\bf B272} (1991) 326.
}
\REF\miyamura{
O.~Miyamura, Nucl.~Phys.~{\bf B}(Proc.~Suppl.){\bf 42} (1995) 538.
}
\REF\shiba{
H.~Shiba and T.~Suzuki, Phys.~Lett.~{\bf B333} (1994) 461.
}
\REF\kitahara{
S.~Kitahara, Y.~Matsubara and T.~Suzuki, Prog.~Theor.~Phys. {\bf 93}
(1995) 1.
}
\REF\ejiri{
S.~Ejiri, S.~Kitahara, Y.~Matsubara and T.~Suzuki,
Phys.~Lett.~{\bf B343} (1995) 304.
}
\REF\origuchi{
O.~Miyamura and S.~Origuchi,
{\it Color Confinement and Hadrons},
(World Scientific, 1995) in press.
}
\REF\woloshyn{
R.~M.~Woloshyn, Phys.~Rev.~{\bf D51} (1995) 6411.
}
\REF\sugamiya{
H.~Suganuma, A.~Tanaka, S.~Sasaki and O.~Miyamura,
Proc.~of~Int.~Symp.~on {\it Lattice Field Theory},
Melbourne, July 11-15, 1995,
to appear in Nucl.~Phys.~{\bf B}.
}
\REF\giacomo{
A.~Di Giacomo, {\it this proceedings}.
}
[\kronfeld-\giacomo].
As 't~Hooft pointed out
[\thooftB],
the nonabelian gauge theory as QCD is reduced to an abelian gauge
theory with magnetic monopoles (QCD-monopoles) by the abelian
gauge fixing, which is defined by the diagonalization of a
gauge-dependent variable $X(x)$.
The QCD-monopole appears from the hedgehog configuration on $X(x)$
\REF\sugaA{
H.~Suganuma, S.~Sasaki and H.~Toki, Nucl.~Phys.~{\bf B435} (1995) 207.
}
\REF\sugaB{
H.~Suganuma, H.~Ichie, S.~Sasaki and H.~Toki,
Prog.~Theor.~Phys.(Suppl.) (1995) in press.
}
\REF\sugaC{
H.~Suganuma, H.~Ichie, S.~Sasaki and H.~Toki,
{\it Color Confinement and Hadrons},
(World Scientific, 1995) in press.
}
[\sugaA-\sugaC]
corresponding to the nontrivial homotopy group
$\pi _2({\rm SU}(N_c)/{\rm U(1)}^{N_c-1})=Z_\infty ^{N_c-1}$,
and its condensation
leads to the color confinement through the dual Meissner effect.
The crucial role of QCD-monopole condensation
to the chiral-symmetry breaking is also supported
by recent lattice studies [\miyamura,\origuchi,\woloshyn]
and the model analyses
\REF\sugaD{
H.~Suganuma, S.~Sasaki and H.~Toki,
{\it Quark Confinement and Hadron Spectrum},
Como, Italy, (World Scientific, 1995) p.238.
\nextline
S.~Sasaki, H.~Suganuma and H.~Toki, {\it ibid} p.241.
}
\REF\sasaki{
S.~Sasaki, H.~Suganuma and H.~Toki, Prog.~Theor.~Phys. {\bf 94} (1995)
373.
}
[\sugaA-\sasaki].
The instanton
\REF\rajaraman{
R.~Rajaraman, {\it Solitons and Instantons},
(North-Holland, Amsterdam, 1982) p.1.
}[\rajaraman]
is also an important topological object in QCD
relating to the U$_{\rm A}$(1) anomaly,
and appears in the Euclidean 4-space ${\bf R}^4$
corresponding to $\pi _3({\rm SU}(N_c))$= $Z_\infty $.

Recent lattice studies [\kronfeld-\giacomo]
indicate the abelian dominance
\REF\iwasaki{
Z.~F.~Ezawa and A.~Iwazaki, Phys.~Rev.~{\bf D25} (1982) 2681.
\nextline
Z.~F.~Ezawa and A.~Iwazaki, Phys.~Rev.~{\bf D26} (1982) 631.
}
[\thooftB,\iwasaki]
for the nonperturbative quantities
in the maximally abelian (MA) gauge and/or in the Polyakov gauge.
{\it
If the system is completely described only by the abelian field,
the instanton would lose the topological basis
for its existence, and therefore it seems unable to
survive in the abelian manifold.
However, even in the abelian gauge, nonabelian components remain
relatively large around the QCD-monopoles, which are nothing
but the topological defects, so that instantons
are expected to survive only around the world lines of
the QCD-monopole in the abelian-dominating system.
}
The close relation between instantons and QCD-monopoles are thus
suggested from the topological consideration.
In this paper, we study the correlation between
instantons and QCD-monopoles both in the lattice theory
\REF\thurner{
S.~Thurner, H.~Markum and W.~Sakuler,
{\it Color Confinement and Hadrons},
(World Scientific, 1995) in press.
}
[\origuchi,\sugamiya,\thurner]
and in the analytical framework
[\sugaC].

\chapter{Analytical Calculation}

\section{Abelian Gauge Fixing and Monopole Charge}

First, the abelian gauge fixing
is studied with attention to the ordering condition [\sugamiya],
which is closely related to the magnetic charge of QCD-monopoles.
In general, the abelian gauge fixing consists of two sequential
procedures.

\item{1.} The diagonalization of a gauge-dependent variable
$X(x)$ by a suitable gauge transformation : $X(x)\rightarrow X_d(x)$
[\sugaA].
The gauge group SU$(N_c)_{\rm local}$ is reduced to
U(1)$^{N_c-1}_{\rm local} \times P^{N_c}_{\rm global}$
by the diagonalization of $X(x)$.

\item{2.} The ordering on the diagonal elements of $X_d(x)$
by imposing the additional condition, for instance,
$$
X_d^1(x) \ge X_d^2(x) \ge ... \ge X_d^{N_c}(x).
\eqn\ONE
$$
The residual gauge group U(1)$^{N_c-1}_{\rm local}
\times P^{N_c}_{\rm global}$
is reduced to U(1)$^{N_c-1}_{\rm local}$
by the ordering condition on $X_d(x)$ [\sugamiya].

The magnetic charge of the QCD-monopole is closely related to
the ordering condition in the diagonalization in the abelian
gauge fixing [\sugamiya].
For instance,
in the SU(2) case,
the hedgehog configuration as $X(x) = ({\bf x} \cdot \tau )$ and
the anti-hedgehog one as $X(x) = -({\bf x} \cdot\tau )$
provide a QCD-monopole with an opposite magnetic charge,
because they are connected by the additional gauge transformation by
$$
\Omega =\exp\{i\pi ({\tau ^1 \over 2} \cos\alpha +{\tau ^2 \over 2} \sin\alpha
)\}
\in P^2_{\rm global}
\eqn\TWO
$$
with an arbitrary constant $\alpha $.
Here, $\Omega $ physically means the rotation of angle $\pi $
in the internal SU(2) space, and it
interchanges the diagonal elements of $X_d(x)$, which leads
a minus sign in the U(1)$_3$ gauge field, $A_\mu ^3(x)$.
Thus, the magnetic charge of the QCD-monopole is settled
by imposing the ordering condition on $X_d(x)$.

$P^{N_c}_{\rm global}$-symmetry is also important for the argument of
gauge dependence.
If a variable holds the residual gauge symmetry in the abelian gauge,
it is proved to be SU($N_c$) gauge invariant [\hioki].
However, one should carefully examine the residual gauge symmetry,
which often includes not only U(1)$^{N_c-1}_{\rm local}$
but also $P^{N_c}_{\rm global}$.
For instance, the dual Ginzburg-Landau theory [\sugaA] is,
strictly speaking, an effective theory
holding U(1)$^{N_c-1}_{\rm local} \times P^{N_c}_{\rm global}$
symmetry.
Hence, gauge dependence of a physical variable
should be carefully checked in terms of the residual gauge symmetry
U(1)$^{N_c-1}_{\rm local} \times P^{N_c}_{\rm global}$
instead of U(1)$^{N_c-1}_{\rm local}$ [\hioki].
As a result, the dual gauge field $\vec B_\mu $ is not
SU($N_c$)-invariant,
because $\vec B_\mu $ is U(1)$^{N_c-1}$-invariant but
is changed under the global $P^{N_c}$ transformation.

\section{
QCD-monopoles in the Polyakov-like Gauge
}

We demonstrate a close relation between instantons
and QCD-monopoles in the Euclidean SU(2) gauge theory in continuum
[\sugaA-\sugaD].
Since there is an ambiguity on the
gauge-dependent variable $X(x)$ to be diagonalized
in the abelian gauge fixing [\thooftB,\sugaA],
it would be a wise way to choose a suitable $X(x)$
so that the instanton configuration can be simply described.
Here, we adopt the Polyakov-like gauge [\sugamiya],
where $A_4(x)$ is diagonalized.
The Polyakov-like gauge has a large similarity to the Polyakov gauge,
because the Polyakov loop $P(x)$ is also diagonal in this gauge.
%

Using the 't~Hooft symbol $\bar \eta ^{a\mu \nu }$,
the multi-instanton solution is written as
[\rajaraman]
$$
A^\mu (x) =i\bar \eta ^{a\mu \nu }{\tau ^a \over 2} \partial^\nu \ln \phi (x)
        =-i{\bar \eta ^{a\mu \nu }\tau ^a \over \phi (x)}
\sum_k {a_k^2 (x-x_k)^\nu  \over |x-x_k|^4},
\hbox{\quad}
\phi (x)  \equiv 1+\sum_k {a_k^2 \over |x-x_k|^2},
\eqn\THREE
$$
where $x_k^\mu  \equiv ({\bf x}_k,t_k)$ and $a_k$ denote the center
coordinate and the size of $k$-th instanton, respectively.
Near the center of $k$-th instanton,
$A_4(x)$ takes a hedgehog configuration around ${\bf x}_k$,
$$
A_4(x) \simeq i {\tau ^a ({\bf x}-{\bf x}_k)^a \over |x-x_k|^2},
\eqn\FOUR
$$
like a single-instanton solution.
In the Polyakov-like gauge,
$A_4(x)$ is diagonalized by a singular gauge transformation,
which provides a QCD-monopole on the center of the hedgehog,
${\bf x} = {\bf x}_k$.
Thus, the center of each instanton is inevitably penetrated
by a QCD-monopole trajectory along the temporal
direction in the Polyakov-like gauge [\sugamiya,\sugaC].
In other words, instantons exist only along the
QCD-monopole trajectories.

\section{QCD-monopole Trajectory in the Multi-Instanton System}

For the single-instanton system,
$A_4(x)$ takes a hedgehog configuration around ${\bf x}_1$,
$$
A_4(x)=-i a_1^2 {\tau ^a ({\bf x}-{\bf x}_1)^a \over
(x-x_1)^2 \cdot \{ (x-x_1)^2+a_1^2 \} }.
\eqn\FIVE
$$
The diagonalization of $A_4(x)$ is carried out using
a time-independent singular gauge transformation
with the gauge function
$$
\Omega ({\bf x})=e^{i\tau _3\phi }\cos {\theta  \over 2}
+{i(\tau _1\cos\alpha +\tau _2\sin\alpha )} \sin {\theta  \over 2}
=
\pmatrix {
e^{i\phi }\cos{\theta  \over 2}     &    ie^{i\alpha }\sin{\theta  \over 2}
\cr
ie^{-i\alpha }\sin{\theta  \over 2}   &    e^{-i\phi }\cos{\theta  \over 2}
}
\eqn\SIX
$$
with $\theta $ and $\phi $ being the polar and azimuthal angles:
${\bf x}-{\bf x}_1=(\sin\theta \cos\phi ,\sin\theta \sin\phi ,\cos\theta )$.
Here, $\alpha $ is an arbitrary constant angle corresponding to the
residual U(1)$_3$ symmetry.
Since $\Omega ({\bf x})$ is time-independent, $A_4(x)$ is simply
transformed as
$$
A_4(x)\rightarrow \Omega ({\bf x})A_4(x)\Omega ^{-1}({\bf x}).
\eqn\SEVEN
$$

After the singular gauge transformation by $\Omega ({\bf x})$,
the abelian gauge field $A_\mu ^3(x)$ has a singular part
stemming from
$$
A_\mu ^{sing}(x)
={1 \over e}\Omega ({\bf x}) \partial_\mu  \Omega ^{-1}({\bf x}),
\eqn\EIGHT
$$
which leads to the QCD-monopole with the magnetic
charge $g=4\pi /e$ [\sugaA].
The QCD-monopole appears at the center of the hedgehog,
${\bf x}={\bf x}_1$, which satisfies $A_4(x)=0$ in Eq.{\FIVE}.
Hence, the QCD-monopole trajectory $x^\mu \equiv({\bf x},t)$
becomes a simple straight line penetrating the center of the instanton
as shown in Fig.1 (a),
$$
{\bf x}={\bf x}_1 \hbox{\quad} (-\infty <t<\infty ),
\eqn\NINE
$$
at the classical level in the Polyakov-like gauge
[\sugaA,\sugamiya,\sugaC].
Similar relation for the QCD-monopole in a single instanton
is found also in the MA gauge
\REF\chernodub{
M.~N.~Chernodub and F.~V.~Gubarev, JETP~Lett. {\bf 62} (1995) 100.
\nextline
A.~Hart, M.~Teper, preprint OUTP-95-44-P (1995), hep-lat/9511016.
}
[\chernodub].

It should be noted that
the singularity of $A_\mu (x)$ at the center of the instanton
can be removed easily by a gauge transformation
to the non-singular gauge [\rajaraman],
where
$$
A_\mu (x)=i {\tau ^a ({\bf x}-{\bf x}_1)^a \over (x^2-x_1^2)+a_1^2}
\eqn\NINEhalf
$$
provides the same QCD-monopole trajectory as mentioned above.
It is also worth mentioning that
the QCD-monopole trajectory
is not changed by the residual U(1)$_3$-gauge transformation,
so that QCD-monopoles in the Polyakov-like gauge
are identical to those, {\it e.g.}, in the temporal gauge: $A_4(x)=0$.

For the single anti-instanton system,
one has only to replace $A_4(x)\rightarrow -A_4(x)$ corresponding to
$\bar \eta ^{a\mu \nu } \rightarrow  \eta ^{a\mu \nu }$ in the above argument
[\rajaraman].
Since this replacement interchanges the hedgehog and the
anti-hedgehog on $A_4(x)$,
it leads to the change of the QCD-monopole charge
as mentioned in Section 2.1.
Then, the QCD-monopole with the opposite magnetic charge, $-g$,
appears and passes through the center of the anti-instanton
as shown in Fig.1 (b).
In Figs.1 (a) and (b), relative difference on the
QCD-monopole charge is expressed by the direction of the arrow.

For the two-instanton system, two instanton centers
can be put on the $zt$-plane by a suitable spatial rotation
in ${\bf R}^3$ without loss of generality,
so that one can set $x_1=y_1=x_2=y_2=0$.
Owing to the axial-symmetry around the $z$-axis of the system,
the QCD-monopole trajectory only appears on the $zt$-plane, and hence
one has only to examine $A_4(x)$ on the $zt$-plane by setting $x=y=0$.
In this case, $A_4(x)$ is already diagonalized on the $zt$-plane:
$$
A_4(z,t;x=y=0)
= -i {\tau^3  \over \phi (z,t)}
\sum_{k=1}^2 a_k^2 {(z-z_k) \over \{ (z-z_k)^2+(t-t_k)^2 \}^2}
\equiv A_4^3(z,t)\tau ^3.
\eqn\TEN
$$
Therefore, the QCD-monopole trajectory $x^\mu =(x,y,z,t)$
is simply given by $x=y=0$ and $A_4^3(z,t)=0$ or
$$
\sum_{k=1}^2 a_k^2 {(z-z_k) \over \{ (z-z_k)^2+(t-t_k)^2 \}^2}=0.
\eqn\ELEVEN
$$
Here, $A_4(x)$ takes a hedgehog or an anti-hedgehog
configuration near the QCD-monopole at each $t$.

We show in Figs. 2 (a),(b) and (c) the typical examples
of the QCD-monopole trajectory in the two-instanton system.
The QCD-monopole trajectories are found to be rather
complicated even at the classical level.
Fig.2 (a) shows the simplest case for two instantons
with the same size, $a_1=a_2$, locating at the same Euclidean time,
$(z_1,t_1)=-(z_2,t_2)=(z_0,0)$.
In this case, the QCD-monopole trajectory
$(z,t)$ is analytically solved [\sugaC] as
$$
z=0
\hbox{\quad or \quad}
t^2=(z_0^2-z^2)+2|z_0|\sqrt{(z_0^2-z^2)},
\eqn\TWELVE
$$
and there appear two junctions and a loop
in the QCD-monopole trajectory [\sugaC].
Here, the QCD-monopole charge calculated
is expressed by the direction of the arrow.

Fig.2 (b) shows an example for two instantons with
the same size, $a_1=a_2$, but a little rotated in ${\bf R}^4$ as
$(z_1,t_1)=-(z_2,t_2)$ =(1,0.05).
In this case, the QCD-monopole trajectory has a folded structure
[\sugamiya].
Fig.2 (c) shows an example for two instantons locating at
the same time $(z_1,t_1)=-(z_2,t_2)$=(1,0), but with a little different
size, $a_2=1.1a_1$.
There appears a QCD-monopole loop in this case [\sugamiya].
Thus, the QCD-monopole trajectories originating from instantons
are very unstable against a small fluctuation relating to
the location or the size of instantons [\sugamiya].

For a general $N$-instanton system with $N \ge 3$,
it is rather difficult to find a suitable gauge transformation
diagonalizing $A_4(x)$, and therefore it is hard to obtain
the QCD-monopole trajectory.
However, the QCD-monopole trajectory can be also
obtained by $x=y=0$ and $A_4(z,t)=0$ as Eq.{\ELEVEN} for
the multi-instantons located on the $zt$-plane.
Hence, we have examined such a special case in the
multi-instanton system.

In general, the QCD-monopole trajectory becomes highly complicated
and unstable in the multi-instanton system even at the classical level,
and a small fluctuation of instantons often changes the topology of
the QCD-monopole trajectory as shown in Fig.2.
In addition, the quantum fluctuation would make it more complicated
and more unstable, which leads to appearance of a long twining
trajectory as a result.
Hence, instantons may contribute to promote monopole condensation,
which is signaled by a long complicated monopole loop in the
lattice QCD simulation [\hioki,\shiba,\kitahara].

\section{QCD-monopole Trajectory in the Thermal-Instanton System}

We also study the thermal instanton system in the Polyakov-like gauge.
The multi-instanton solution at finite temperature $T$ is
given by
$$
\eqalign{
A^\mu (x)
&=i\bar \eta ^{a\mu \nu }{\tau ^a \over 2} \partial^\nu \ln \phi (x)
=i\bar \eta ^{a\mu \nu }{\tau ^a \over 2} \partial^\nu \phi (x)/\phi (x),
\cr
\phi (x)&=1+\sum_k a_k^2 \sum_{n=-\infty }^\infty
{1 \over ({\bf x}-{\bf x}_k)^2+(t-t_k-n/T)^2} \cr
&=1+\pi T \sum_k {a_k^2 \over |{\bf x}-{\bf x}_k|}
\cdot
{{\rm sinh}(2\pi T|{\bf x}-{\bf x}_k|) \over
{\rm cosh}(2\pi T|{\bf x}-{\bf x}_k|) -\cos \{2\pi T(t-t_k)\}}.
}
\eqn\THIRTEEN
$$
In this system, $A_4(x)$ is given by
$$
\eqalign{
A_4(x)
=&-i{\pi T \tau ^a \over 2\phi }\sum_k
{a_k^2 ({\bf x}-{\bf x}_k)^a \over |{\bf x}-{\bf x}_k|^3}
\bigl(
{{\rm sinh}(2\pi T|{\bf x}-{\bf x}_k|) \over
{\rm cosh}(2\pi T|{\bf x}-{\bf x}_k|) -\cos \{2\pi T(t-t_k)\}} \cr
& -2\pi T |{\bf x}-{\bf x}_k| \cdot
{1-{\rm cosh}(2\pi T|{\bf x}-{\bf x}_k|) \cos\{2\pi T(t-t_k) \} \over
[{\rm cosh}(2\pi T|{\bf x}-{\bf x}_k|) -\cos \{2\pi T(t-t_k)\}]^2}
\bigr)
}
\eqn\FOURTEEN
$$

At the high-temperature limit $T\rightarrow \infty $,
$$
A_4(x) \simeq -{i\pi T \over 2\phi } \tau^a
\sum_k {a_k^2 ({\bf x}-{\bf x}_k)^a \over |{\bf x}-{\bf x}_k|^3}
\eqn\SIXTEEN
$$
becomes time-independent, so that $A_4({\bf x})$ can be diagonalized
using a time-independent gauge transformation by $\hat \Omega ({\bf x})$,
$$
A_4({\bf x})\rightarrow  \hat \Omega ({\bf x})A_4({\bf x}) \hat \Omega
^{-1}({\bf x})
=A_4^d({\bf x}),
\eqn\FIFTEEN
$$
where QCD-monopoles appear at
the points ${\bf x}_s$ satisfying $A_4({\bf x}_s)=0$,
These points ${\bf x}_s$ includes
all the centers of instantons, ${\bf x}_k$,
and become the centers of the (anti-) hedgehog configuration
on $A_4({\bf x})$.
Thus, the QCD-monopole trajectory
is reduced to simple straight lines
$$
{\bf x}={\bf x}_s \hbox{\quad} (-\infty <t<\infty ),
\eqn\SEVENTEEN
$$
where each instantons are penetrated in the temporal direction.
Such a simplification of the QCD-monopole trajectory
may corresponds to the deconfinement phase transition through
the vanishing of QCD-monopole condensation
\REF\ichie{
H.~Ichie, H.~Suganuma and H.~Toki, Phys.~Rev.~{\bf D52} (1995) 2944.
}
[\kronfeld-\ejiri,\sugaB,\ichie].

For the thermal two-instanton system, all instanton centers
can be put on the $zt$-plane by a suitable spatial
rotation in ${\bf R}^3$
like the two-instanton system at $T=0$,
so that one can set as $x_k=y_k=0$ $(k=1,2)$.
Owing to the axial-symmetry around the $z$-axis of the system,
the QCD-monopole trajectory only appears on the $zt$-plane,
where $A_4(x)$ in Eq.{\FOURTEEN} is already diagonalized.
Hence, the QCD-monopole trajectory $x^\mu =(x,y,z,t)$
is simply given by $x=y=0$ and $A_4(z,t;x=y=0)=0$.
Here, $A_4(x)$ takes a hedgehog or an anti-hedgehog
configuration near the QCD-monopole at each $t$.

We show in Fig.3 the typical examples of the QCD-monopole trajectory
in the thermal two-instanton system.
As temperature goes high, the trajectory tends to be
straight lines in the temporal direction.
There also appears the QCD-monopole with the opposite
magnetic charge at the point satisfying $A_4(x)=0$.
The topology of the QCD-monopole trajectory is drastically changed
at $T_c \simeq 0.6 d^{-1}$, where $d$ is the distance between the
two instantons. If one adopts $d \sim 1{\rm fm}$
as a typical mean distance between instantons, such a topological
change occurs at $T_c \sim 120 {\rm MeV}$ [\sugamiya].

\chapter{Instanton and Monopole on Lattice}

\section{Framework}

We study the correlation between instantons and QCD-monopoles
in the maximally abelian (MA) gauge [\origuchi] and
in the Polyakov gauge using the SU(2) lattice with
$16^4$ and $\beta =2.4$.
All measurements are done every 500 sweeps after a
thermalization of 1000 sweeps using the heat-bath algorithm.
After generating the gauge configurations,
we examine the monopole dominance
[\kronfeld-\giacomo]
for the topological charge using the following procedure
[\origuchi,\sugamiya].

\item{\rm 1.}
The abelian gauge fixing is carried out by diagonalizing
$R(s)=\sum_{\mu }U_\mu (s)\tau ^3U_\mu ^{-1}(s)$ in the MA gauge,
and/or the Polyakov loop $P(s)$ in the Polyakov gauge.

\item{\rm 2.} The SU(2) link variable $U_\mu (s)$ is factorized into
the abelian link variable $u_\mu (s)=\exp\{i\tau _3\theta _\mu (s)\}$
and the `off-diagonal' factor $M_\mu (s)$ [\suzuki-\sugamiya],
$$
U_\mu (s)
=\pmatrix{
\sqrt{1-|c_\mu (s)|^2} & -c_\mu (s)       \cr
c_\mu ^*(s)            & \sqrt{1-|c_\mu (s)|^2}
}
\pmatrix{
e^{i\theta _\mu (s)} &  0       \cr
0             &  e^{-i\theta _\mu (s)}
}
\equiv
M_\mu (s)u_\mu (s),
\eqn\EIGHTEEN
$$
where $c_\mu (s)$ transforms as the charged matter field.

\item{\rm 3.} The abelian field strength
$\theta _{\mu \nu }\equiv \partial_\mu \theta _\nu -\partial_\nu \theta _\mu $
is decomposed as
$$
\theta _{\mu \nu }(s)=\bar \theta _{\mu \nu }(s)+2\pi M_{\mu \nu }(s)
\eqn\TWENTYONE
$$
with $-\pi <\bar \theta _{\mu \nu }(s)<\pi $ and $M_{\mu \nu }(s) \in {\bf Z}$.
Here, $\bar \theta _{\mu \nu }(s)$ and $2\pi M_{\mu \nu }(s)$ correspond to
the regular photon part and the Dirac string part, respectively
\REF\DGT{
T.~DeGrand and D.~Toussaint, Phys.~Rev.~{\bf D22} (1980) 2478.
}
[\DGT].

\item{\rm 4.} The U(1) gauge field $\theta _\mu (s)$ is decomposed
as $\theta _\mu (s)=\theta ^{Ph}_\mu (s)+\theta ^{Ds}_\mu (s)$
[\kronfeld-\sugamiya,\DGT],
where the regular part $\theta ^{Ph}_\mu (s)$ and
the singular part $\theta ^{Ds}_\mu (s)$ are obtained from
$\bar \theta _{\mu \nu }(s)$ and $2\pi M_{\mu \nu }(s)$, respectively,
$$
\theta _\mu ^{Ds}(s)=2\pi \sum_{s'} G(s-s')\partial^\lambda M_{\lambda \mu
}(s'),
\hbox{\quad}
\theta _\mu ^{Ph}(s)=\sum_{s'} G(s-s')\bar \partial^\lambda \bar \theta
_{\lambda \mu }(s'),
\eqn\TWENTYTWO
$$
using the lattice Coulomb propagator $G(s)$ in the Landau gauge,
which satisfies $\partial^2_s G(s-s')=\delta ^4(s-s')$.
The singular part carries almost the same amount of
the magnetic current as the original U(1) field, whereas
it scarcely carries the electric current.
The situation is just the opposite in the regular part.
For this reason, we regard the singular part as `monopole-dominating',
and the regular part as `photon-dominating' [\origuchi,\sugamiya].

\item{\rm 5.} The corresponding SU(2) variables are reconstructed
from $\theta ^{Ph}_\mu (s)$ and $\theta ^{Ds}_\mu (s)$
by multiplying the off-diagonal factor $M_\mu (s)$:
$$
U^{Ds}_\mu (s)=M_\mu (s)\exp\{i\tau _3\theta _\mu ^{Ds}(s)\}.
\hbox{\quad}
U^{Ph}_\mu (s)=M_\mu (s)\exp\{i\tau _3\theta _\mu ^{Ph}(s)\},
\eqn\TWENTYTHREE
$$

\item{\rm 6.}
By using $U_\mu (s)$, $U^{Ph}_\mu (s)$ and $U^{Ds}_\mu (s)$,
we calculate the topological charge $Q$ and the integral $I_Q$
of the absolute value of the topological density,
$$
Q={1 \over 16\pi ^2}\int d^4x \Tr(G_{\mu \nu }\tilde G_{\mu \nu }),
\hbox{\quad}
I_Q \equiv {1 \over 16\pi ^2}\int d^4x |\Tr(G_{\mu \nu }\tilde G_{\mu \nu })|,
\eqn\TWENTYFOUR
$$
and the action divided by $8\pi ^2$,
$$
S={1 \over 16\pi ^2}\int d^4x \Tr(G_{\mu \nu }G_{\mu \nu }).
\eqn\TWENTYFIVE
$$
Here, $I_Q$ is introduced to get information on
the instanton and anti-instanton pair [\origuchi,\sugamiya].
Three sets of quantities are thus obtained:
$$
\eqalign{
U&_\mu (s) \ \rightarrow \{Q({\rm SU(2)}), I_Q({\rm SU(2)}), S({\rm SU(2)})\}
\cr
U&_\mu ^{Ds}(s)\rightarrow \{Q({\rm Ds}),    I_Q({\rm Ds}),    S({\rm Ds})\}
\cr
U&_\mu ^{Ph}(s)\rightarrow \{Q({\rm Ph}),    I_Q({\rm Ph}),    S({\rm Ph})\}
}
\eqn\TWENTYSIX
$$
Of course,
$\{Q({\rm SU(2)})$, $I_Q({\rm SU(2)})$, $S({\rm SU(2)})\}$
defined with the full SU(2) link variable is a set of the
ordinary quantities.
On the other hand,
$\{Q({\rm Ds})$, $I_Q({\rm Ds})$, $S({\rm Ds})\}$ and
$\{Q({\rm Ph})$, $I_Q({\rm Ph})$, $S({\rm Ph})\}$
provide the information on the singular (monopole-dominating) part
and the regular (photon-dominating) part, respectively.

\item{\rm7.} The correlations among these quantities are
examined using the Cabibbo-Marinari cooling method.

We prepare 40 samples for the MA gauge and the Polyakov gauge,
respectively.
These simulations have been performed on the Intel Paragon
XP/S(56node) at the Institute for Numerical Simulations
and Applied Mathematics of Hiroshima University.
Since quite similar results have been obtained in the MA gauge
[\origuchi] and in the Polyakov gauge,
only latter case is shown below.

\section{Monopole Dominance for Topological Charge on Lattice}

Fig.4 shows the correlation among $Q({\rm SU(2)})$,
$Q({\rm Ds})$ and $Q({\rm Ph})$
at various cooling sweeps in the Polyakov gauge.
A strong correlation is found between
$Q({\rm SU(2)})$ and $Q({\rm Ds})$, which is defined in singular
(monopole-dominating) part. Such a strong correlation
remains even at 80 cooling sweeps.
On the other hand, $Q({\rm Ph})$ quickly vanishes only by several
cooling sweeps, and no correlation is seen between $Q({\rm Ph})$
and $Q({\rm SU(2)})$.

We show in Fig.5 the cooling curves for $Q$, $I_Q$ and $S$
in a typical example with $Q({\rm SU(2)}) \ne 0$
in the Polyakov gauge.
Similar to the full SU(2) case,
$Q({\rm Ds})$, $I_Q({\rm Ds})$ and $S({\rm Ds})$ in the singular
(monopole-dominating) part tends to remain finite
during the cooling process.
On the other hand, $Q({\rm Ph})$, $I_Q({\rm Ph})$
and $S({\rm Ph})$ in the regular part quickly vanish by
only several cooling sweeps.
Therefore, instantons seems unable to live in the regular
(photon-dominating) part, but only survive in the
singular (monopole-dominating) part in the abelian gauges.

We show in Fig.6
the cooling curves for $Q$, $I_Q$ and $S$ are examined
in the case with  $Q({\rm SU(2)})=0$ in the Polyakov gauge.
Similar to the full SU(2) results,
$I_Q({\rm Ds})$ and $S({\rm Ds})$ decrease slowly and
remain finite even at 70 cooling sweeps,
which indicates the existence of the instanton and
anti-instanton pair in the singular (monopole-dominating) part.
On the other hand, $I_Q({\rm Ph})$ and $S({\rm Ph})$ quickly
vanish, which indicates the absence of such a topological pair
excitation in the regular (photon-dominating) part [\sugamiya].

In conclusion, {\it the monopole dominance for the topological charge
is found both in the MA gauge and in the Polyakov gauge.}
In particular, instantons would survive only in the singular
(monopole-dominating) part in the abelian gauges,
which agrees with the result in our previous analytical study.

\chapter{Summary and Concluding Remarks}

We have studied the relation between instantons and monopoles
in the abelian gauge.
Simple topological consideration indicates that instantons
survive only around the QCD-monopole trajectory, which is
the topological defect.

We have found a close relation between instantons
and the QCD-monopole trajectory
in the Polyakov-like gauge, where $A_4(x)$ is to be diagonalized.
Every instantons are penetrated by the world lines of QCD-monopoles
inevitably.
The QCD-monopole trajectory in ${\bf R}^4$
tends to be folded and complicated
in the multi-instanton system,
although it becomes a simple straight line
in the single-instanton solution.
The QCD-monopole trajectory is very unstable against a small
fluctuation on the location and the size of instantons.

We have also studied the thermal instanton system in the
Polyakov-like gauge.
At the high-temperature limit, the QCD-monopole trajectory
becomes straight lines in the temporal direction.
The QCD-monopole trajectory drastically changes its topology
at a high temperature.

We have studied the correlation between instantons and QCD-monopoles.
in the maximally abelian (MA) gauge and/or the Polyakov gauge
on SU(2) lattice with $16^4$ ($\beta $=2.4).
The abelian link variable $u_\mu (s)$ is decomposed into
the singular (monopole-dominating) part $u_\mu ^{Ds}(s)$ and the regular
(photon-dominating) part $u_\mu ^{Ph}(s)$.
We have measured the instanton numbers, $Q({\rm Ds})$ and $Q({\rm Ph})$,
using the SU(2) variables, $U_\mu ^{Ds}(s)$ and $U_\mu ^{Ph}(s)$, which
are reconstructed by multiplying the off-diagonal matter factor to
$u_\mu ^{Ds}(s)$ and $u_\mu ^{Ph}(s)$, respectively.
Topological charge $Q({\rm Ds})$ in the singular (monopole-dominating)
part remains to have a finite number during the cooling process.
On the other hand, $Q({\rm Ph})$ quickly vanishes by several
cooling sweeps, which indicates the absence of instantons in
the regular (photon-dominating) part.
Thus, instantons cannot live in the regular (photon-dominating)
part, but survive in the singular (monopole-dominating) part.
We have found a strong correlation between $Q({\rm Ds})$
and the ordinary topological charge $Q({\rm SU(2)})$
during the cooling process.
We have found such a monopole dominance for the topological
charge both in the MA gauge and in the Polyakov gauge.

\refout

\centerline{Figure Captions}

\item{Fig.1}
The QCD-monopole trajectory (a) in the single-instanton system,
(b) in the single anti-instanton system.
The (anti-)instanton is denoted by a small circle.

\item{Fig.2} Examples of the QCD-monopole trajectory
in the two-instanton system with
(a) $(z_1,t_1)=-(z_2,t_2)$ =(1,0), $a_1=a_2$;
(b) $(z_1,t_1)=-(z_2,t_2)$ =(1,0.05), $a_1=a_2$;
(c) $(z_1,t_1)=-(z_2,t_2)=(1,0), a_2=1.1a_1$.

\item{Fig.3}
The QCD-monopole trajectory in the thermal two-instanton
system
with $(z_1,t_1)=-(z_2,t_2)=$ $d/2\cdot(1,0)$
and $a_1=a_2$ (a) at $T^{-1}=2d$; (b) at $T^{-1}=1.5d$.
The same with $(z_1,t_1)=-(z_2,t_2)=d/2\cdot(1,0.05)$
and $a_1=a_2$ (c) at $T^{-1}=2d$; (d) at $T^{-1}=1.5d$.

\item{Fig.4} (a) Correlations between
$Q$(Ds) and $Q$(SU(2)) at various cooling sweeps.
(b)  Correlations between
$Q$(Ph) and $Q$(SU(2)) at various cooling sweeps.

\item{Fig.5} Cooling curves for
(a) $Q$(SU(2)), $I_Q$(SU(2)), $S$(SU(2));
(b) $Q$(Ds), $I_Q$(Ds), $S$(Ds);
(c) $Q$(Ph), $I_Q$(Ph), $S$(Ph)
in the case with  $Q({\rm SU(2)}) \ne 0.$

\item{Fig.6} Cooling curves for
(a) $Q$(SU(2)), $I_Q$(SU(2)), $S$(SU(2));
(b) $Q$(Ds), $I_Q$(Ds), $S$(Ds);
(c) $Q$(Ph), $I_Q$(Ph), $S$(Ph)
in the case with  $Q({\rm SU(2)}) = 0$.

\end